\documentclass[letterpaper,aps,prd,superscriptaddress,showpacs,nofootinbib,floatfix,twocolumn]{revtex4}

\usepackage{amssymb}
\usepackage{graphicx}

\newcommand{\gevc}{\mbox{GeV/$c$}}
\newcommand{\gevcsq}{\mbox{GeV/$c^{2}$}}
\newcommand{\pt}{\mbox{$p_T$}~}
\newcommand{\jpsi}{\mbox{$J/\psi~$}}

\newcommand{\costheta}{\mbox{$\cos \theta^{*}~$}}
\newcommand{\ee}{\mbox{$e^{+}e^{-}$~}}
\newcommand{\full}{\mbox{$\sqrt{s}=$ 200 GeV~}}
\newcommand{\pp}{\mbox{$p$+$p$~}}
\newcommand{\cc}{\mbox{$c\bar{c}~$}}

\newcommand{\qq}{\mbox{$Q\bar{Q}~$}}
\newcommand{\Op}{\mbox{$\mathcal{O}$}}
\def\mean#1{\ensuremath{\left<#1\right>}}

\def\mean#1{\mbox{$\left<#1\right>$}}

\def\func#1{\left(#1\right)}
\def\kept#1{\left|#1\right>}

\begin{document}

\title{Transverse momentum dependence of $J/\psi$~polarization \\
       at midrapidity in $p+p$ collisions at $\sqrt{s}=200$~GeV.}

\newcommand{\abilene}{Abilene Christian University, Abilene, Texas 79699, USA}
\newcommand{\acadsin}{Institute of Physics, Academia Sinica, Taipei 11529, Taiwan}
\newcommand{\banaras}{Department of Physics, Banaras Hindu University, Varanasi 221005, India}
\newcommand{\barc}{Bhabha Atomic Research Centre, Bombay 400 085, India}
\newcommand{\bnlcoll}{Collider-Accelerator Department, Brookhaven National Laboratory, Upton, New York 11973-5000, USA}
\newcommand{\bnlphys}{Physics Department, Brookhaven National Laboratory, Upton, New York 11973-5000, USA}
\newcommand{\caucr}{University of California - Riverside, Riverside, California 92521, USA}
\newcommand{\charlesczech}{Charles University, Ovocn\'{y} trh 5, Praha 1, 116 36, Prague, Czech Republic}
\newcommand{\ciae}{China Institute of Atomic Energy (CIAE), Beijing, People's Republic of China}
\newcommand{\cns}{Center for Nuclear Study, Graduate School of Science, University of Tokyo, 7-3-1 Hongo, Bunkyo, Tokyo 113-0033, Japan}
\newcommand{\colorado}{University of Colorado, Boulder, Colorado 80309, USA}
\newcommand{\columbia}{Columbia University, New York, New York 10027 and Nevis Laboratories, Irvington, NY 10533, USA}
\newcommand{\czechtech}{Czech Technical University, Zikova 4, 166 36 Prague 6, Czech Republic}
\newcommand{\dapnia}{Dapnia, CEA Saclay, F-91191, Gif-sur-Yvette, France}
\newcommand{\debrecen}{Debrecen University, H-4010 Debrecen, Egyetem t{\'e}r 1, Hungary}
\newcommand{\elte}{ELTE, E{\"o}tv{\"o}s Lor{\'a}nd University, H - 1117 Budapest, P{\'a}zm{\'a}ny P. s. 1/A, Hungary}
\newcommand{\fit}{Florida Institute of Technology, Melbourne, Florida 32901, USA}
\newcommand{\fsu}{Florida State University, Tallahassee, Florida 32306, USA}
\newcommand{\gsu}{Georgia State University, Atlanta, Georgia 30303, USA}
\newcommand{\hiroshima}{Hiroshima University, Kagamiyama, Higashi-Hiroshima 739-8526, Japan}
\newcommand{\ihepprot}{IHEP Protvino, State Research Center of Russian Federation, Institute for High Energy Physics, Protvino, 142281, Russia}
\newcommand{\illuiuc}{University of Illinois at Urbana-Champaign, Urbana, Illinois 61801, USA}
\newcommand{\instpasczech}{Institute of Physics, Academy of Sciences of the Czech Republic, Na Slovance 2, 182 21 Prague 8, Czech Republic}
\newcommand{\isu}{Iowa State University, Ames, Iowa 50011, USA}
\newcommand{\jinrdubna}{Joint Institute for Nuclear Research, 141980 Dubna, Moscow Region, Russia}
\newcommand{\kek}{KEK, High Energy Accelerator Research Organization, Tsukuba, Ibaraki 305-0801, Japan}
\newcommand{\kfki}{KFKI Research Institute for Particle and Nuclear Physics of the Hungarian Academy of Sciences (MTA KFKI RMKI), H-1525 Budapest 114, POBox 49, Budapest, Hungary}
\newcommand{\korea}{Korea University, Seoul, 136-701, Korea}
\newcommand{\kurchatov}{Russian Research Center ``Kurchatov Institute", Moscow, Russia}
\newcommand{\kyoto}{Kyoto University, Kyoto 606-8502, Japan}
\newcommand{\labllr}{Laboratoire Leprince-Ringuet, Ecole Polytechnique, CNRS-IN2P3, Route de Saclay, F-91128, Palaiseau, France}
\newcommand{\lawllnl}{Lawrence Livermore National Laboratory, Livermore, California 94550, USA}
\newcommand{\losalamos}{Los Alamos National Laboratory, Los Alamos, New Mexico 87545, USA}
\newcommand{\lpc}{LPC, Universit{\'e} Blaise Pascal, CNRS-IN2P3, Clermont-Fd, 63177 Aubiere Cedex, France}
\newcommand{\lund}{Department of Physics, Lund University, Box 118, SE-221 00 Lund, Sweden}
\newcommand{\mass}{Department of Physics, University of Massachusetts, Amherst, Massachusetts 01003-9337, USA }
\newcommand{\muenster}{Institut f\"ur Kernphysik, University of Muenster, D-48149 Muenster, Germany}
\newcommand{\muhlenberg}{Muhlenberg College, Allentown, Pennsylvania 18104-5586, USA}
\newcommand{\myongji}{Myongji University, Yongin, Kyonggido 449-728, Korea}
\newcommand{\nagasaki}{Nagasaki Institute of Applied Science, Nagasaki-shi, Nagasaki 851-0193, Japan}
\newcommand{\newmex}{University of New Mexico, Albuquerque, New Mexico 87131, USA }
\newcommand{\nmsu}{New Mexico State University, Las Cruces, New Mexico 88003, USA}
\newcommand{\ornl}{Oak Ridge National Laboratory, Oak Ridge, Tennessee 37831, USA}
\newcommand{\orsay}{IPN-Orsay, Universite Paris Sud, CNRS-IN2P3, BP1, F-91406, Orsay, France}
\newcommand{\peking}{Peking University, Beijing, People's Republic of China}
\newcommand{\pnpi}{PNPI, Petersburg Nuclear Physics Institute, Gatchina, Leningrad region, 188300, Russia}
\newcommand{\riken}{RIKEN Nishina Center for Accelerator-Based Science, Wako, Saitama 351-0198, JAPAN}
\newcommand{\rikjrbrc}{RIKEN BNL Research Center, Brookhaven National Laboratory, Upton, New York 11973-5000, USA}
\newcommand{\rikkyo}{Physics Department, Rikkyo University, 3-34-1 Nishi-Ikebukuro, Toshima, Tokyo 171-8501, Japan}
\newcommand{\saispbstu}{Saint Petersburg State Polytechnic University, St.  Petersburg, Russia}
\newcommand{\saopaulo}{Universidade de S{\~a}o Paulo, Instituto de F\'{\i}sica, Caixa Postal 66318, S{\~a}o Paulo CEP05315-970, Brazil}
\newcommand{\seoulnat}{System Electronics Laboratory, Seoul National University, Seoul, Korea}
\newcommand{\stonybrkc}{Chemistry Department, Stony Brook University, Stony Brook, SUNY, New York 11794-3400, USA}
\newcommand{\stonycrkp}{Department of Physics and Astronomy, Stony Brook University, SUNY, Stony Brook, New York 11794, USA}
\newcommand{\subatech}{SUBATECH (Ecole des Mines de Nantes, CNRS-IN2P3, Universit{\'e} de Nantes) BP 20722 - 44307, Nantes, France}
\newcommand{\tenn}{University of Tennessee, Knoxville, Tennessee 37996, USA}
\newcommand{\titech}{Department of Physics, Tokyo Institute of Technology, Oh-okayama, Meguro, Tokyo 152-8551, Japan}
\newcommand{\tsukuba}{Institute of Physics, University of Tsukuba, Tsukuba, Ibaraki 305, Japan}
\newcommand{\vandy}{Vanderbilt University, Nashville, Tennessee 37235, USA}
\newcommand{\waseda}{Waseda University, Advanced Research Institute for Science and Engineering, 17 Kikui-cho, Shinjuku-ku, Tokyo 162-0044, Japan}
\newcommand{\weizmann}{Weizmann Institute, Rehovot 76100, Israel}
\newcommand{\yonsei}{Yonsei University, IPAP, Seoul 120-749, Korea}
\affiliation{\abilene}
\affiliation{\acadsin}
\affiliation{\banaras}
\affiliation{\barc}
\affiliation{\bnlcoll}
\affiliation{\bnlphys}
\affiliation{\caucr}
\affiliation{\charlesczech}
\affiliation{\ciae}
\affiliation{\cns}
\affiliation{\colorado}
\affiliation{\columbia}
\affiliation{\czechtech}
\affiliation{\dapnia}
\affiliation{\debrecen}
\affiliation{\elte}
\affiliation{\fit}
\affiliation{\fsu}
\affiliation{\gsu}
\affiliation{\hiroshima}
\affiliation{\ihepprot}
\affiliation{\illuiuc}
\affiliation{\instpasczech}
\affiliation{\isu}
\affiliation{\jinrdubna}
\affiliation{\kek}
\affiliation{\kfki}
\affiliation{\korea}
\affiliation{\kurchatov}
\affiliation{\kyoto}
\affiliation{\labllr}
\affiliation{\lawllnl}
\affiliation{\losalamos}
\affiliation{\lpc}
\affiliation{\lund}
\affiliation{\mass}
\affiliation{\muenster}
\affiliation{\muhlenberg}
\affiliation{\myongji}
\affiliation{\nagasaki}
\affiliation{\newmex}
\affiliation{\nmsu}
\affiliation{\ornl}
\affiliation{\orsay}
\affiliation{\peking}
\affiliation{\pnpi}
\affiliation{\riken}
\affiliation{\rikjrbrc}
\affiliation{\rikkyo}
\affiliation{\saispbstu}
\affiliation{\saopaulo}
\affiliation{\seoulnat}
\affiliation{\stonybrkc}
\affiliation{\stonycrkp}
\affiliation{\subatech}
\affiliation{\tenn}
\affiliation{\titech}
\affiliation{\tsukuba}
\affiliation{\vandy}
\affiliation{\waseda}
\affiliation{\weizmann}
\affiliation{\yonsei}
\author{A.~Adare} \affiliation{\colorado}
\author{S.~Afanasiev} \affiliation{\jinrdubna}
\author{C.~Aidala} \affiliation{\mass}
\author{N.N.~Ajitanand} \affiliation{\stonybrkc}
\author{Y.~Akiba} \affiliation{\riken} \affiliation{\rikjrbrc}
\author{H.~Al-Bataineh} \affiliation{\nmsu}
\author{J.~Alexander} \affiliation{\stonybrkc}
\author{K.~Aoki} \affiliation{\kyoto} \affiliation{\riken}
\author{L.~Aphecetche} \affiliation{\subatech}
\author{J.~Asai} \affiliation{\riken}
% ...
\collaboration{PHENIX Collaboration} \noaffiliation

\begin{abstract}

We report the measurement of the transverse momentum dependence
of inclusive $J/\psi$ polarization in $p+p$ collisions at
$\sqrt{s}=$ 200 GeV performed by the PHENIX Experiment at RHIC.
The $J/\psi$ polarization is studied in the helicity,
Gottfried-Jackson, and Collins-Soper frames for $p_T<5$ GeV/$c$
and $|y|<0.35$.  The polarization in the helicity and
Gottfried-Jackson frames is consistent with zero for all
transverse momenta, with a slight (1.8 sigma) trend towards
longitudinal polarization for transverse momenta above
2~GeV/$c$.  No conclusion is allowed due to the limited
acceptance in the Collins-Soper frame and the uncertainties of
the current data.  The results are compared to observations for
other collision systems and center of mass energies and to
different quarkonia production models.

\end{abstract}


\maketitle


\section{Introduction}
\label{section:introduction}

\setcounter{section}{1}\setcounter{equation}{0}

Quarkonia production in high-energy hadronic collisions is an 
essential tool for investigating QCD.  The $Q\bar{Q}$ pair is 
produced in a hard scattering involving gluons, which is followed by 
a hadronization process that forms the bound state.  These formation 
and hadronization steps are the subject of many studies.  Initial 
tests of quarkonia production models using \jpsi cross sections 
measurements are still inconclusive~\cite{Adare:2006kf}, suggesting 
that other observables would be useful to challenge the different 
production models.  For example, a key piece of information to help 
pin down the mechanism of heavy quarkonia (\cc and $b\bar{b}$) 
production and the bound state formation is the angular distribution 
of its decay leptons.

The angular distribution of spin-$\frac{1}{2}$ lepton decay from 
quarkonium (spin 1) is derived from density matrix elements of the 
production amplitude and parity conservation 
rules~\cite{PhysRevD.18.2447,Gottfried:1964nx,Collins:1977iv}.  The 
angular distribution integrated over the azimuthal angle is given by
\begin{equation}
\label{eq:costheta}
\frac{d\sigma}{d \cos\theta^{*}} = A\left( {1}+{\lambda
  \cos^{2}\theta^{*}}\right),
\end{equation}
\noindent where $A$ is a normalization factor and $\theta^{*}$ is the 
angle between the momentum vector of one lepton in the polarization 
quarkonium rest frame and the longitudinal direction ($\hat{z}$ 
coordinate) of a selected polarization vector (frame).  The 
polarization parameter $\lambda$ is related to the diagonal elements 
of the density matrix of the production amplitude and contains both 
the longitudinal $\left(\sigma_L\right)$ and transverse 
$\left(\sigma_T\right)$ components of the quarkonium cross section
\begin{equation}
  \label{eq:lambda}
  \lambda = \frac{\sigma_{T} - 2\sigma_{L}}{\sigma_{T} + 2\sigma_{L}}.
\end{equation}
\noindent The quarkonium polarization is longitudinal (transverse) in 
a given frame if $\lambda$ is negative (positive).

\begin{figure}
\centering
  \includegraphics[width=0.9\linewidth]{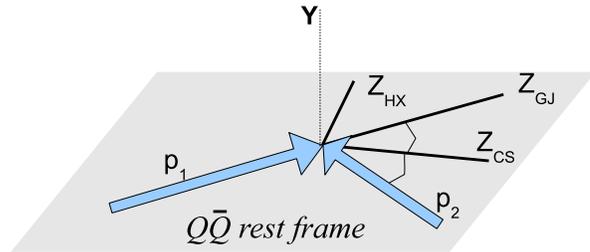}
  \caption{\label{fig:frames}Definition of the polarization frames:
    helicity(HX), Gottfried-Jackson(GJ) and Collins-Soper(CS) frames.}
\end{figure}

The most common polarization frame used in analyses performed at 
collider experiments is where $\hat{z}$ is the quarkonium momentum.  
Polarization measured in this manner is referred to as being in the 
helicity frame (HX)~\cite{PhysRevD.18.2447}.  In fixed target 
experiments the most frequently used polarization frame has $\hat{z}$ 
as one of the colliding hadrons momentum in the quarkonium rest 
frame, namely, the Gottfried-Jackson frame 
(GJ)~\cite{Gottfried:1964nx}.  Another polarization frame, used 
primarily for the studies of Drell-Yan production, is the 
Collins-Soper frame (CS)~\cite{Collins:1977iv} that defines $\hat{z}$ 
as the bisector between the directions of the first colliding parton 
and of the opposite of the second colliding parton in the dilepton 
rest frame.  A diagram representing the three polarization frames is 
shown in Fig.~\ref{fig:frames}.  The amplitude and the sign of 
$\lambda$ depend on the frame used in the measurement.  The natural 
polarization axis for the production process can be defined as that 
where the lepton decay azimuthal angle distribution is symmetric and 
$\lambda$ is maximum \cite{Faccioli:2008dx}.  In such a frame, the 
density matrix of the production amplitude is diagonal.

Several quarkonium production models have been proposed to describe 
the perturbative terms which are relevant for \qq production, while 
other models include non-perturbative terms related to the formation 
of the bound state.  The various models predict different 
polarizations and are described below.

In the Color Evaporation Model (CEM) \cite{Fritzsch:1977ay}, 
quarkonia production is assumed to be a fixed fraction of the pQCD 
cross section for invariant masses between twice the mass of the 
heavy quark ($c$ or $b$) and twice the mass of the open heavy quark 
meson ($D$ or $B$).  This model has reasonable agreement with most of 
the measured quarkonia cross sections but no predictive power for the 
polarization \cite{Ramona:2009}.  Nevertheless, according to 
\cite{Amundson:1996qr}, multiple soft gluon exchanges destroy the 
polarization of the heavy quark pair.

The earliest Color Singlet Model (CSM) was a calculation of the
leading order \mbox{$gg \rightarrow$ S-wave charmonium $ +~g$}
process where the relative momentum of the $Q\bar{Q}$ pair with
respect to the quark mass $m_Q$ is neglected and the pair is produced
on-shell \cite{Baier:1981uk,Brambilla:2004wf,Lansberg:2006dh}.  The
$Q\bar{Q}$ binding is calculated from potential model wave functions.
\jpsi yield measurements reported by CDF \cite{Abe:1997jz} and PHENIX
\cite{Adare:2006kf} are largely underestimated by this model.  The
\jpsi polarization predicted by LO CSM is transverse in the HX frame
\cite{Chung:2009xr}.  Subsequent calculations also included NLO terms
\cite{Campbell:2007ws,Artoisenet:2007xi,Gong:2008sn}, NNLO terms
\cite{Artoisenet:2008fc,Lansberg:2008gk} and an s-channel cut
contribution that allows off-shell $c\bar{c}$ quarks to end up in the
bound state \cite{Haberzettl:2007kj}.  These calculations show large
changes in the yield and polarization relative to the earlier
calculations.  The new calculations of the \jpsi yield is closer to
what is observed in PHENIX and CDF for $p_T<10~ \gevc$.  The \jpsi
polarization is predominantly longitudinal in the HX frame according
to these new calculations.

Non-relativistic QCD (NRQCD) effective theory \cite{Bodwin:1994jh} 
makes use of short distance $\left(m_Q \right)$ and non-relativistic 
$\left(m_Q\nu^2\right)$ terms, where $\nu$ is the typical quark 
velocity in the quarkonium rest frame.  A typical $\nu$ for charm 
(bottom) is $0.3c~(0.1c)$.  The S-wave charmonium is described as a 
series of intermediate color singlet$^{(1)}$ or color octet$^{(8)}$ 
state contributions
\begin{eqnarray}
 \kept{\psi_{Q}} &=& \Op(1)\kept{^3S_1^{(1)}} +
 \Op(\nu)\kept{^3P_J^{(8)}g}\\ \nonumber
 &+& \Op\func{\nu^2}\kept{^3S_1^{(8)}gg} +
\Op\func{\nu^2}\kept{^3S_0^{(8)}g} + \cdots,
\end{eqnarray}
\noindent where the spectroscopic notation $^{2S+1}L_J$ is used.  
The non-perturbative operators $\Op\func{\nu}$ are parametrized using 
experimental results.  Since the singlet state has a small 
contribution to the yield, this model is also referred as the Color 
Octet Model (COM).  Using constraints from the CDF cross section 
\jpsi data, reasonable agreement is obtained with PHENIX yield 
results assuming \jpsi production is dominated by gluon fusion in the 
$^1S_0^{(8)}$ and $^3P_0^{(8)}$ intermediate states for $\pt< 
5~\gevc$ \cite{Cooper:2004qe,Chung:2009xr}.  Calculations performed 
in \cite{Beneke:1996tk} estimated \mbox{$\lambda(^1S_0^{(8)})=0$} and 
\mbox{$\lambda(^3P_0^{(8)})=-0.05$} indicating a very small 
longitudinal polarization from direct \jpsi in this \pt range.  
Numerical estimations \cite{Beneke:1996yw,Braaten:1999qk} and 
subsequent NLO corrections \cite{Gong2009197} supports that the 
polarization for $p_T \gg M_{J/\psi}$, where production from gluon 
fragmentation is supposed to be important, is predominantly 
transverse in the HX frame.

The \jpsi polarization in hadronic collisions was studied in fixed 
target experiments at $\sqrt{s} \leq$ 39 GeV 
\cite{PhysRevLett.42.948,PhysRevLett.45.2092,Biino:1987qu,Gribushin:1995rt,Tzamarias:1990ij,Abt:2009nu,Alexopoulos:1997yd,Chang:2003rz}.  
These experiments predominantly covered $|x_F|>0$ and 
\mbox{$p_T<5~$\gevc}.  In \cite{Faccioli:2008dx} it was noted that 
\jpsi polarization measured in the CS frame by HERA-B 
\cite{Abt:2009nu} and by E866/NuSea \cite{Chang:2003rz} smoothly 
changes from longitudinal to transverse with the total momentum.  
The polarization observed in CDF at \mbox{$\sqrt{s}=1.96~$TeV} in 
midrapidity for \mbox{$p_T>5~$ \gevc}$~$ showed a small longitudinal 
polarization in the HX frame \cite{Abulencia:2007us}.  This result 
contradicts the first LO CSM and COM expectations.

Complimentary \jpsi polarization measurements in \pp collisions at 
\full can help elucidate the production mechanism.  Moreover, it is 
expected that the polarization of \jpsi is modified in the presence 
of nuclear matter effects in $d$+Au collisions and hot and dense 
matter in Au+Au collisions \cite{Ioffe:2003rd}.  Thus future 
measurements of \jpsi polarization in $d$+Au and Au+Au at RHIC 
demands a good reference from \pp collisions.

This paper reports the transverse momentum dependence of the \jpsi 
polarization for $|y|<0.35$ in the HX, GJ and CS reference frames.  
The study was performed in the dielectron decay channel for 
$p_T<5~\gevc~$.  The experimental apparatus used to measure electron 
decays from \jpsi mesons is detailed in section \ref{sec:apparatus}.  
The procedure followed to obtain the \costheta distributions, the 
corresponding polarization parameters and their uncertainties are 
explained in \mbox{section \ref{sec:analysis}}.  The results, 
comparison with measurements at other facilities and interpretation 
in the context of current theoretical models are presented in section 
\ref{sec:results}.

\section{Experimental apparatus and \jpsi identification}
\label{sec:apparatus}

This analysis was performed with data collected in \pp collisions at 
\full during the 2006 RHIC Run with the PHENIX central arm detectors 
\cite{Adler:2003zu}.  The geometrical coverage for single electrons 
corresponds to pseudorapidity $|\eta|<$0.35.  Each one of the 
roughly back-to-back two arms covers $\Delta\phi = \pi/2$.

Data were recorded using a minimum-bias trigger that required at 
least one hit in each of the two beam-beam counters (BBC) located at 
$3.0 <|\eta|< 3.9$ and scanning approximately 50\% of the \pp cross 
section.  A dedicated trigger (EMCal RICH Trigger - ERT) was also 
used to select events with at least one electron candidate.  The ERT 
required a minimum energy in any 2$\times$2 group of the 
Electromagnetic Calorimeter (EMCal) towers~\footnote{Corresponding to 
$\Delta \eta \times \Delta \phi = 0.02 \times 0.02$ rad} and 
associated hits in the Ring Imaging $\check{C}$erenkov detector 
(RICH) in coincidence with the minimum-bias trigger condition.  The 
EMCal energy threshold was set to 0.4 GeV and \mbox{0.6 GeV} for two 
different periods during the data taking run.

Collisions within $\pm$30 cm of the center of the detector along the 
beam direction were used in this analysis.  After data quality 
selection, the number of collisions sampled was 143 billion BBC 
triggers, corresponding to an integrated luminosity of $\int 
\mathcal{L}=(6.2\pm0.6)~$ pb$^{-1}$.

Electron candidates were selected from tracks reconstructed in the 
Drift Chamber (DCH) and in the Pad Chamber (PC) with momentum larger 
than 0.5 \gevc.  Electron identification was achieved by requiring 
the tracks to be associated with at least one fired phototube within 
a ring radius 3.4 cm $<R_{ring}<$ 8.4 cm centered on the projected 
track position in the RICH.  In addition, the presence of a matching 
energy cluster in the EMCal was required within four sigma in both 
the position and expected energy/momentum ratio.  Since the hadronic 
background in the \jpsi mass region is small in \pp collisions, only 
loose electron identification criteria were used.

\begin{figure}
  \centering
  \includegraphics[width=0.9\linewidth]{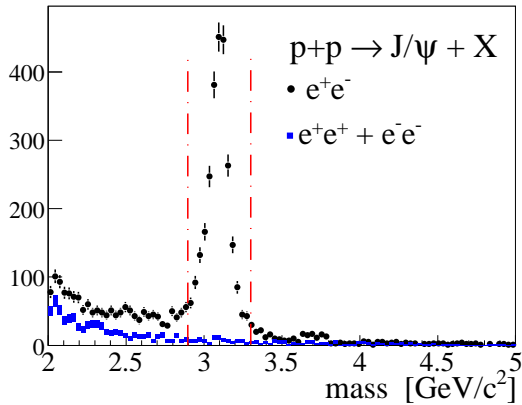}
  \caption{\label{fig:jpsi_peak}Invariant mass of dielectrons in the
    \jpsi mass range.  Dashed lines represents the mass range used in
    the polarization analysis.}
\end{figure}

Dielectron pairs from \jpsi decays were counted in the invariant mass 
range \mbox{$\in[2.9,3.2]$ \gevcsq}.  The combinatorial background was 
estimated using like-sign $\left(e^+e^+~\textrm{and}~ e^-e^-\right)$ 
pairs.  Since we evaluated the ERT efficiency using \jpsi simulation, we 
required that the ERT segment was fired by one of the \jpsi decayed 
electrons.  Hence, only pairs with at least one electron matching 
geometrically the position of an actual ERT trigger in the event were 
accepted.  The dielectron mass distribution in the \jpsi mass region is 
shown in Figure~\ref{fig:jpsi_peak}.  The signal/(combinatorial background) 
ratio was 28.  After combinatorial background subtraction, we counted 
$2442~\pm~51~e^+e^-$ pairs with $p_T<5~\gevc~$ in the selected \jpsi mass 
range.  These counts include a residual continuum background, which 
consists mainly of correlated open heavy quark decays to electrons.  This 
background was found to be less than 10\%.

\begin{figure}
\centering
  \hspace{-0.5cm}
   \includegraphics[width=1.0\linewidth]{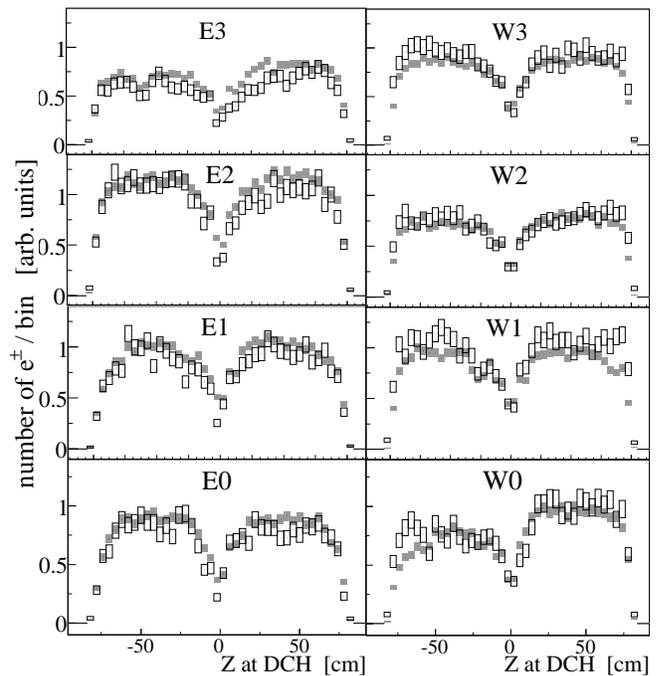}
  \caption{\label{fig:acceptance_zed} Distribution of single
  electrons versus the $Z$ coordinate of the track in the drift chamber
  for real (open boxes) and simulated (closed boxes) data for
  different sectors (0 is the bottom and 3 is the top ones) in the
  east (left) and west (right) detector arms.  The box-height for each
  point corresponds to its statistical uncertainty.}
\end{figure}

\section{Analysis Procedure}
\label{sec:analysis}

The angular dependence of the detector response for electrons from 
\jpsi decays was estimated using a full GEANT3 \cite{GEANT} based 
detector simulation.  Dead channels or malfunctioning regions were 
removed from the detector simulation and from the real data analysis.  
The experimental acceptance was checked by simulating single 
electrons with collision $z$ vertex and \pt distributions weighted to 
reproduce observed distributions of electron candidates from real 
data.  The simulated detector acceptance for these single electrons 
was compared to that for real data for different azimuthal sectors 
(see Fig.~\ref{fig:acceptance_zed}).  Remaining differences in the 
acceptance between simulated and real data were attributed to 
conversions $\gamma \rightarrow \ee$ in the detector support 
structure at large $z$ which were not included in the simulation.  
These differences were accounted for in the systematic uncertainty 
listed in Table~\ref{tab:typeB_errors}.

\begin{table*}
\caption{\label{tab:typeB_errors}Systematic uncertainties in the \pt
  dependent polarization measurement in the helicity and
  Gottfried-Jackson (in parentheses) frames.}
\begin{ruledtabular}
\begin{tabular}{lcccc} 
  Description &
  $[0,1]~\gevc$ &
  $[1,2]~\gevc$ &
  $[2,5]~\gevc$ &
  $[0,5]~\gevc$\\ \hline 
  Acceptance &
  0.006 (0.036) &
  0.006 (0.012) &
  0.006 (0.008) &
  0.006 (0.024)\\ 
  Polarization bias in acceptance &
  0.022 (0.047) &
  0.0011 (0.005) &
  0.008 (0.031) &
  0.012 (0.032)\\ 
  Continuum fraction &
  $^{+0.033}_{-0.021}~\left(^{+0.091}_{-0.014}\right)$ &
  $^{+0.023}_{-0.027}~\left(^{+0.032}_{-0.062}\right)$ &
  $^{+0.014}_{-0.039}~\left(^{+0.023}_{-0.070}\right)$ &
  $^{+0.019}_{-0.026}~\left(^{+0.032}_{-0.058}\right)$\\ 
  Input \pt in simulation &
  0.034 (0.062) &
  0.005 (0.049) &
  0.024 (0.028) &
  0.034 (0.054)\\ 
  Input $y$, Z vertex in simulation  & 
  0.000 (0.007) &
  0.000 (0.007) &
  0.000 (0.007) &
  0.000 (0.007) \\ 
  Run-by-run fluctuations &
  0.019 (0.123) &
  0.016 (0.035) &
  0.016 (0.020) &
  0.017 (0.050)\\ 
  ERT efficiency &
  0.017 (0.110) &
  0.015 (0.051) &
  0.018 (0.024) &
  0.015 (0.043)\\ \hline
  TOTAL &
  $^{+0.06}_{-0.05}~\left(^{+0.21}_{-0.19}\right)$ &
  $^{+0.03}_{-0.04}~\left(^{+0.09}_{-0.10}\right)$ &
  $0.04~\left(^{+0.06}_{-0.09}\right)$ &
  $0.05~\left(^{+0.09}_{-0.11}\right)$\\ 
\end{tabular}
\end{ruledtabular}
\end{table*}

\begin{figure} 
\centering
  \includegraphics[width=1.0\linewidth]{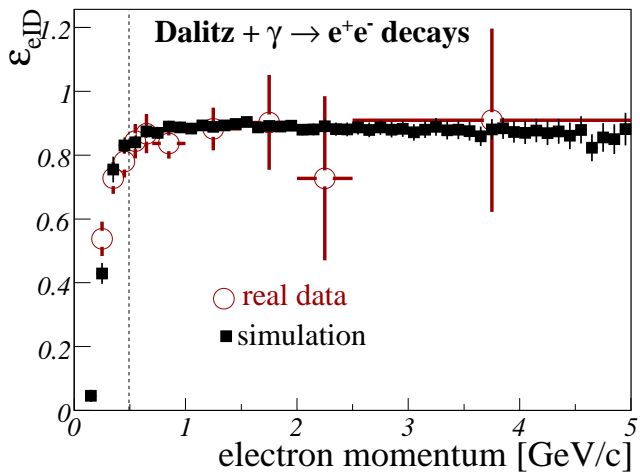}
  \caption{\label{fig:eid_eff} Single electron identification
    efficiency estimated using full reconstructed Dalitz decays from
    real data (open circles) and simulation (full squares).  Dotted
    line represents the minimum \pt for the electron used to
    reconstruct \jpsi decays.}
\end{figure}

The detector response to electrons in the simulation was tuned to 
match the data.  A clean sample of electrons in the data was obtained 
by selecting electrons from Dalitz decays and photon conversions in 
the beam pipe which were identified by their very low invariant mass 
\cite{:2009qk}.  Figure~\ref{fig:eid_eff} shows the comparison 
between the momentum dependent electron identification efficiency 
($\varepsilon_{eID}$) for single electrons from fully reconstructed 
Dalitz and photon conversion decays in minimum-bias data and 
simulation.  Good agreement above 0.5 \gevc$~$ was achieved within 
the statistical uncertainties.

The \pt dependence of the ERT efficiency was estimated for each one 
of the 8 EMCal sectors by taking the fraction of minimum-bias single 
electron candidates that fired the ERT.  These efficiencies were used 
in the ERT simulation.  Changes in the trigger thresholds and channel 
masks in the ERT during the run period were used in the simulation in 
order to reproduce realistic run conditions.

The tuned detector simulation was used to reproduce the measurement 
of \jpsi dielectron pairs and to match their momentum, rapidity and 
vertex distributions.  The kinematics of the simulated \jpsi were 
estimated in four steps:

\begin{enumerate}

\item Unpolarized \jpsi \ee pairs were generated with uniform 
distributions in rapidity $\left(|y|<0.5\right)$, 
\mbox{$p_T~\left(p_T<7~\frac{GeV}{c} \right)$}, azimuthal angle 
$\left(-\pi<\phi<\pi\right)$, and collision vertex along the beam 
axis $Z$ \mbox{$\left(|Z_{vertex}|<40~ \textrm{cm}\right)$}.

\item The \jpsi \pt distribution obtained after applying the 
efficiency and acceptance corrections agrees with the previous result 
\cite{Adare:2006kf}.  A Kaplan function 
\mbox{$\frac{d\sigma}{dydp_T}=\frac{A 
p_T}{\left[1+\left(p_T/b\right)^2\right]^{n}}$} was fit to the \pt 
distribution (Fig.~\ref{fig:jpsi_fit}), and a Gaussian function was 
fit to the rapidity dependence of the \jpsi yield reported in 
\cite{Adare:2006kf} and to the collision $Z$ vertex distribution.

\begin{figure}
\centering
  \includegraphics[width=1.0\linewidth]{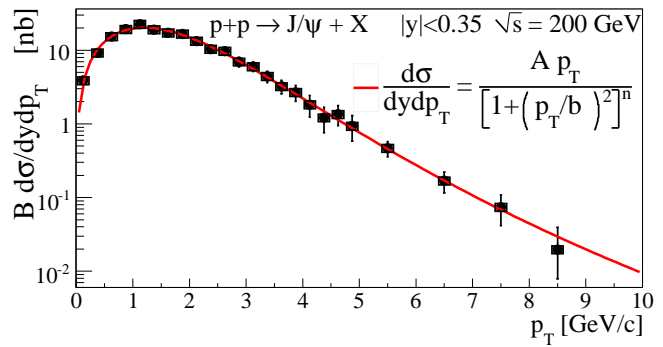}
  \caption{\label{fig:jpsi_fit} Fit to \jpsi yield times dielectron
branching ratio ($B$) after detector acceptance and efficiency
corrections for the real data with 
$A=28.7 \pm 1.0$~nb/\gevc, $b=3.41 \pm 0.21$ \gevc, and $n=4.6 \pm 0.4$.
}
\end{figure}

\begin{figure*}[tbh]
    \includegraphics[width=0.326\linewidth]{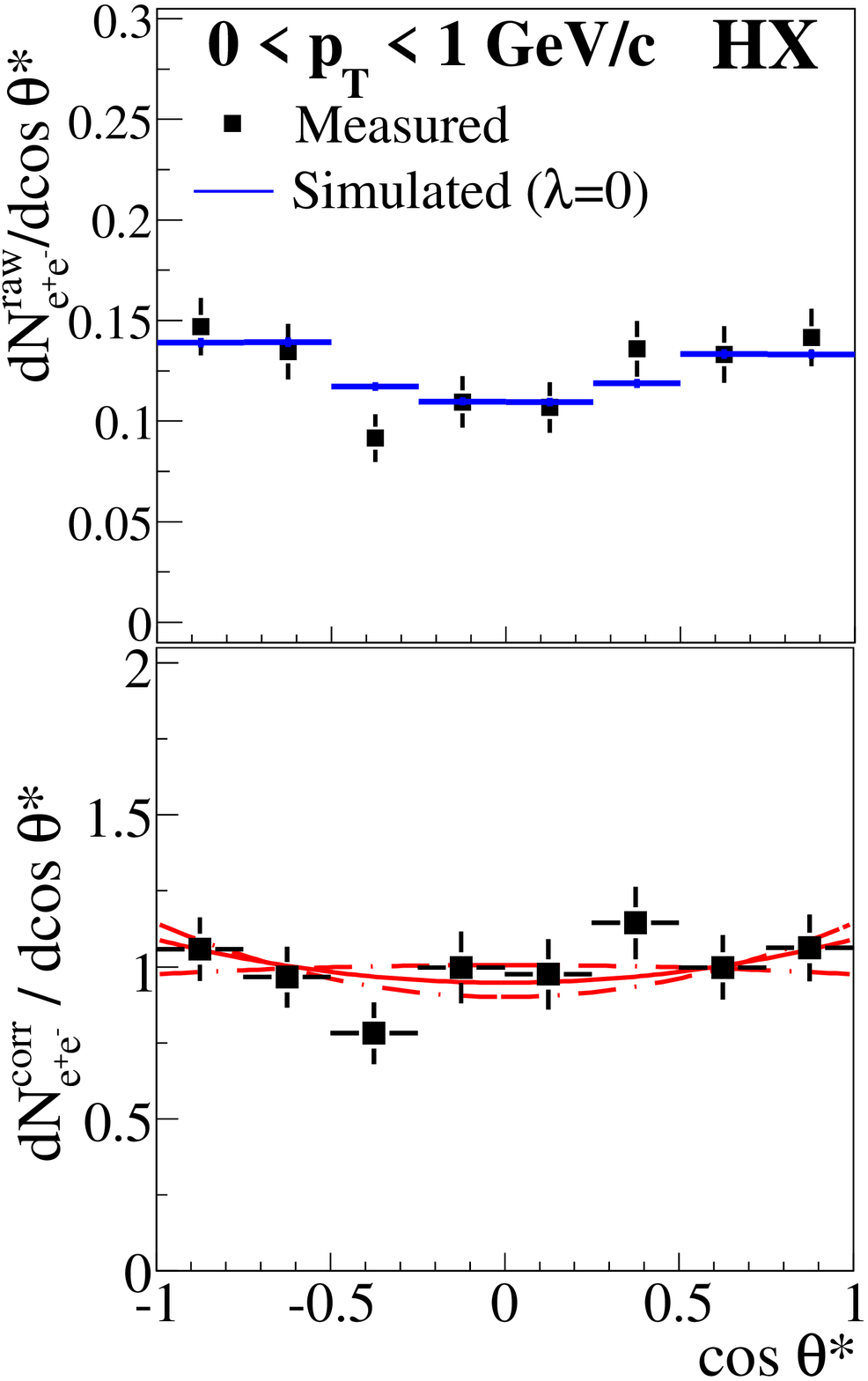}
    \includegraphics[width=0.288\linewidth]{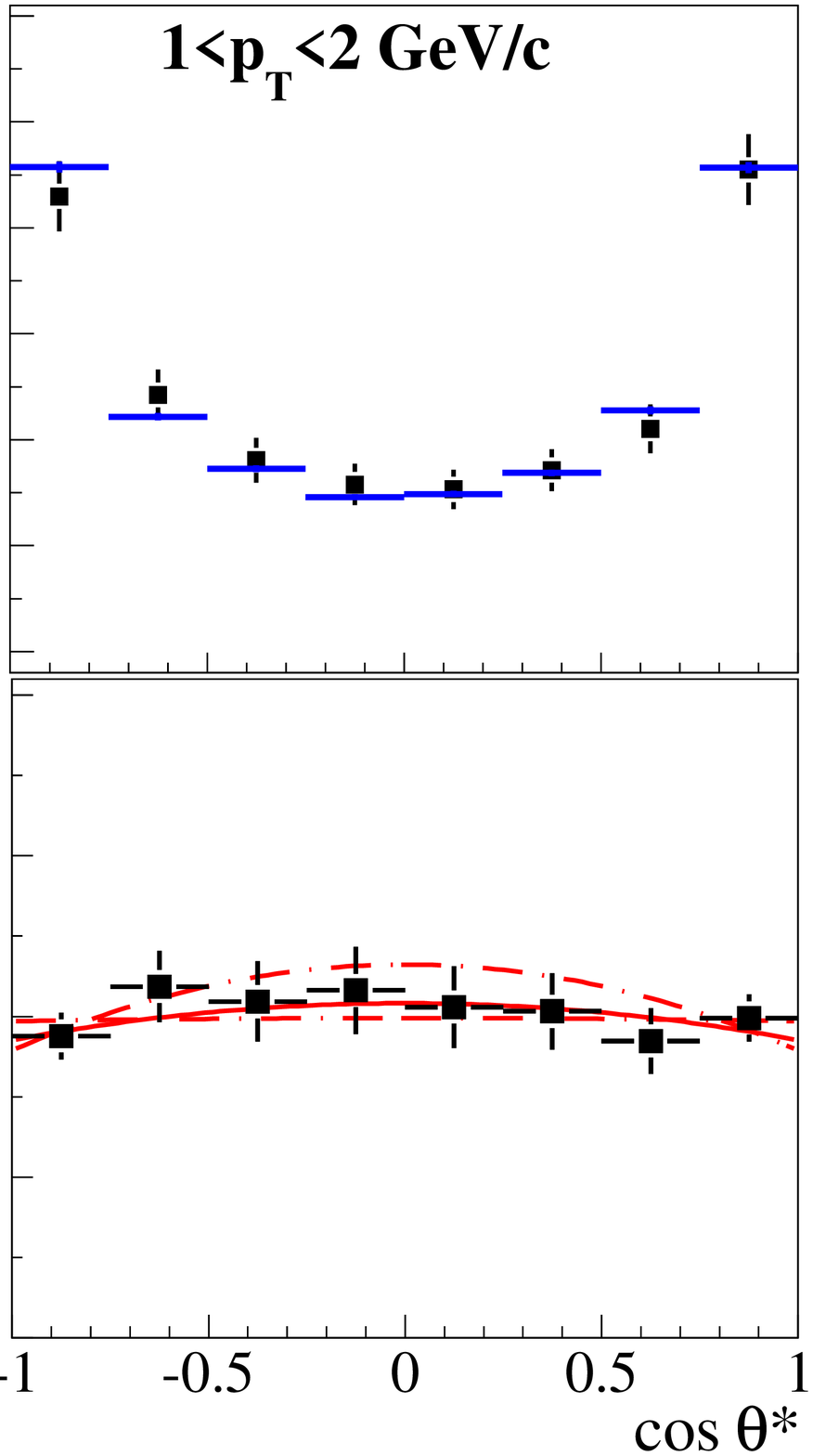}
    \includegraphics[width=0.288\linewidth]{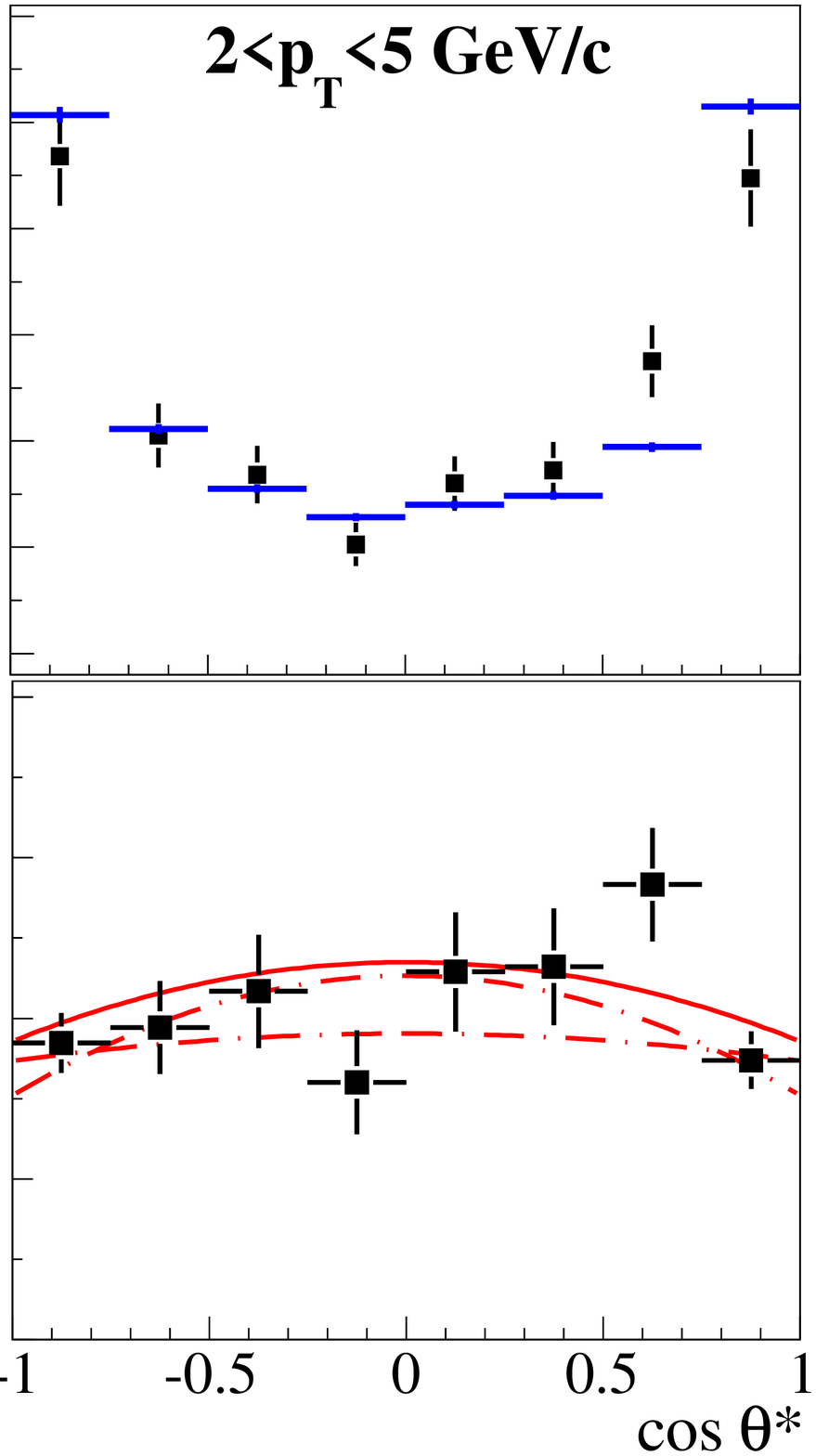}
    \includegraphics[width=0.326\linewidth]{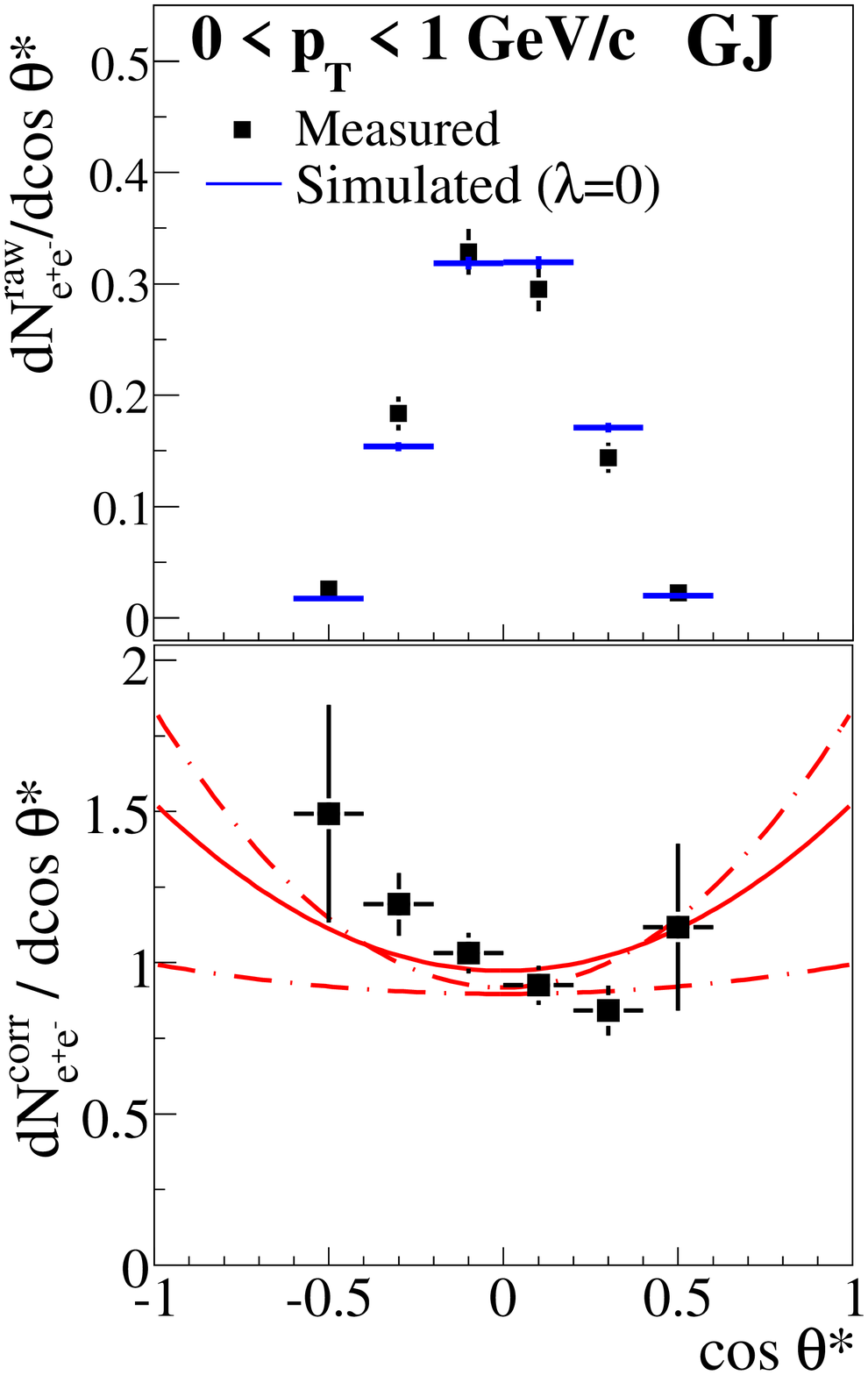}
    \includegraphics[width=0.288\linewidth]{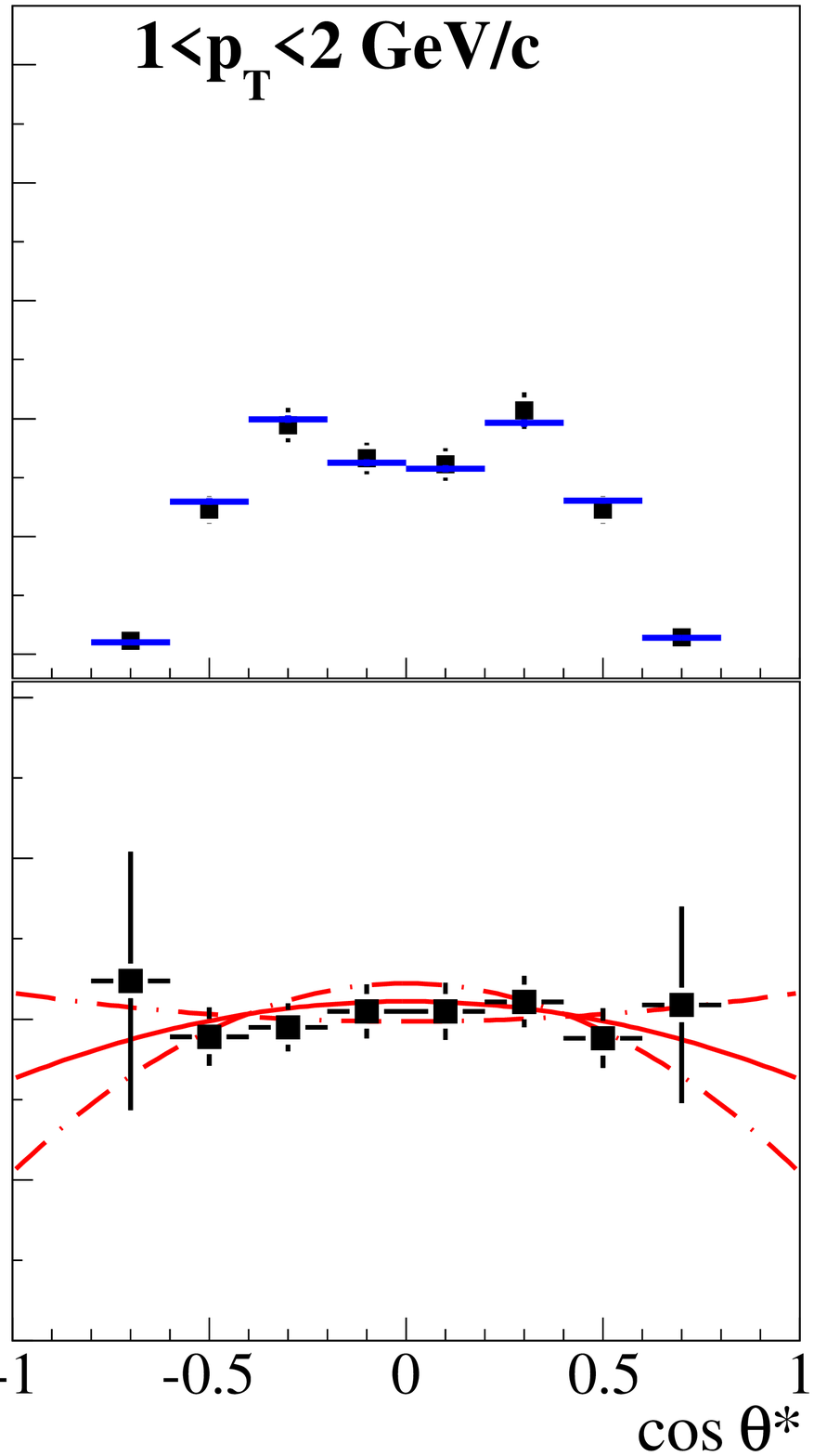}
    \includegraphics[width=0.288\linewidth]{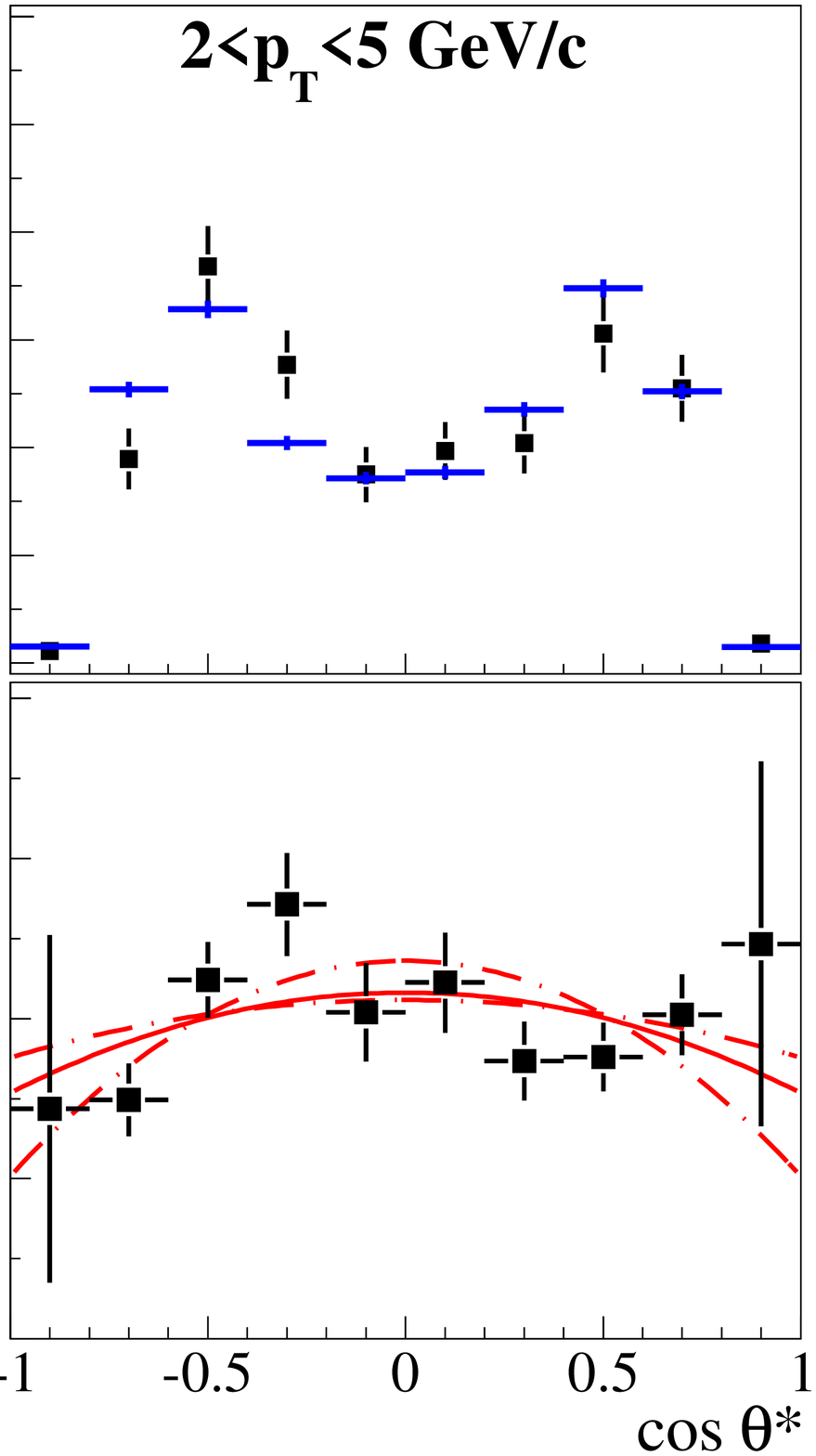}
\caption{\label{fig:costheta_HX_GJ} 
Each of the six plots shows (upper half) \costheta distributions of 
positrons decayed from (solid points) measured and (horizontal bars) 
simulated \jpsi mesons and (lower half) acceptance corrected 
distributions obtained from the ratio between real and simulated 
\jpsi distributions.  Fits to Eqn. 1 are represented as solid lines.
Dashed lines correspond to one standard variation of the parameters 
in the fit.  The top three plots are for the helicity (HX) frame 
and the bottom three are for the Gottfried-Jackson (GJ) frame, 
where the smaller~\costheta range causes larger uncertainties 
of the fits.}
\end{figure*}

\item The fitted \pt, rapidity and collision vertex functions were 
then used to re-weight the simulated \jpsi events.  The top half of 
each plot in Figs.~\ref{fig:costheta_HX_GJ}~and~\ref{fig:costheta_CS} 
shows the \costheta distributions in the HX, GJ, and CS frames of \ee 
pairs in the \jpsi mass range obtained in \jpsi simulation and real 
data \footnote{The \costheta resolution estimated in the simulation 
was 0.08 in the HX, 0.025 in the GJ and 0.007 in the CS frames.  
These resolutions are much smaller than the bin width of the 
\costheta distributions used in the polarization analysis.} after 
combinatorial background subtraction.  The simulated and real data 
distributions are functions of the detector acceptance and efficiency 
and the original $dN_{e^+e^-}/d\costheta$ in the \jpsi mass range.  
The bottom panels show the ratio between the real data and simulated 
\mbox{$\lambda=0$} distributions, corresponding to the acceptance 
corrected \costheta distributions.

Equation~(\ref{eq:costheta}) was fitted to these acceptance corrected 
\costheta distributions with no constraints on the parameters.  
Solid lines are the most likely fits and dashed lines represent 68\% 
confidence level interval.  In the CS frame, the fit returned a 
polarization which was out of the physical limits $\left(\lambda 
\in[-1,1]\right)$.  This was a result of the small acceptance for the 
\costheta distribution in the PHENIX central arms for this frame, 
leading to a large statistical uncertainty on its polarization 
measurements.  Thus, the CS frame is no longer considered in this 
article.

\item Any asymmetry in the electron decay distribution, i.e.  
$\lambda \neq 0$, can change the detector acceptance.  Hence, the 
fourth and final step of the simulation was to apply a weight in 
\costheta to the simulated \jpsi by using the $\lambda$ obtained in 
the third step.  When using this realistic angular distribution for 
the \pt dependent acceptance, and the corresponding uncertainties, we 
obtained a variation in the yield up to $\pm$8\% for \mbox{$p_T < 
5~$\gevc} that corresponds to changes in polarization results no 
larger than 0.02 in the HX frame and 0.05 in the GJ frame.  These 
variations were accounted for in the systematic uncertainties.

\end{enumerate}

\begin{figure}
\centering
    \includegraphics[width=0.7\linewidth]{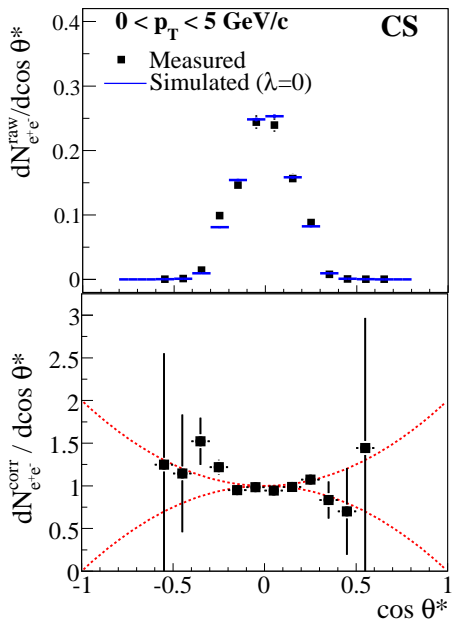}
  \caption{\label{fig:costheta_CS}(top) Same as in
    As in Fig.~\protect\ref{fig:costheta_HX_GJ}, but now for the 
    Collins-Soper frame.  Dotted lines on bottom panel correspond to 
    Eq.~(\protect\ref{eq:costheta}) for $\lambda=\pm1$.}
\end{figure}

We also estimated the contribution to the \jpsi po\-la\-ri\-za\-tion 
from the continuum background by measuring $\lambda$ in the 
dielectron mass range $[1.7,2.3]~$\gevcsq.  The acceptance and 
efficiency corrections were performed using simulated $D\bar{D} 
\rightarrow \ee$ decays, the dominant source of \ee pairs in 
$[1.7,2.3]~$\gevcsq, according to the analysis in \cite{:2008asa}.  
The polarization in this mass range is consistent with zero, with 
values between $\pm 0.3$ in the HX and $\pm 0.9$ in the GJ frame.  
The 10\% continuum contribution can change the measured polarization 
in the \jpsi mass range by at most $^{+0.05}_{-0.02}$ in HX frame and 
$^{+0.17}_{-0.14}$ in GJ frame and was included in the systematic 
uncertainties.

The $\lambda$ measurement is also sensitive to differences between 
acceptance in simulated and in real data, run-by-run condition 
variations, uncertainties in rapidity, $Z$ vertex, and transverse 
momentum shape inputs to the simulation, as well as the ERT 
efficiency \pt shape.  These uncertainties were introduced as 
variations in the efficiency and weighting parameters for different 
detector sectors in the simulation.  Resulting variations in 
$\lambda$ were accounted for as systematic uncertainties and are 
listed in Table \ref{tab:typeB_errors}.  The systematic uncertainties 
are correlated between different \pt ranges.  The total systematic 
uncertainty is taken to be the quadratic sum of these components, 
assuming they are uncorrelated.  Additional checks included the
variation of the minimum momentum requirement of the single electrons
and the rejection of tracks going to the edges of the detector. These
variations returned only statistical fluctuations in the polarization
results.

\section{Results}
\label{sec:results}

\begin{figure}
    \includegraphics[width=0.9\linewidth]{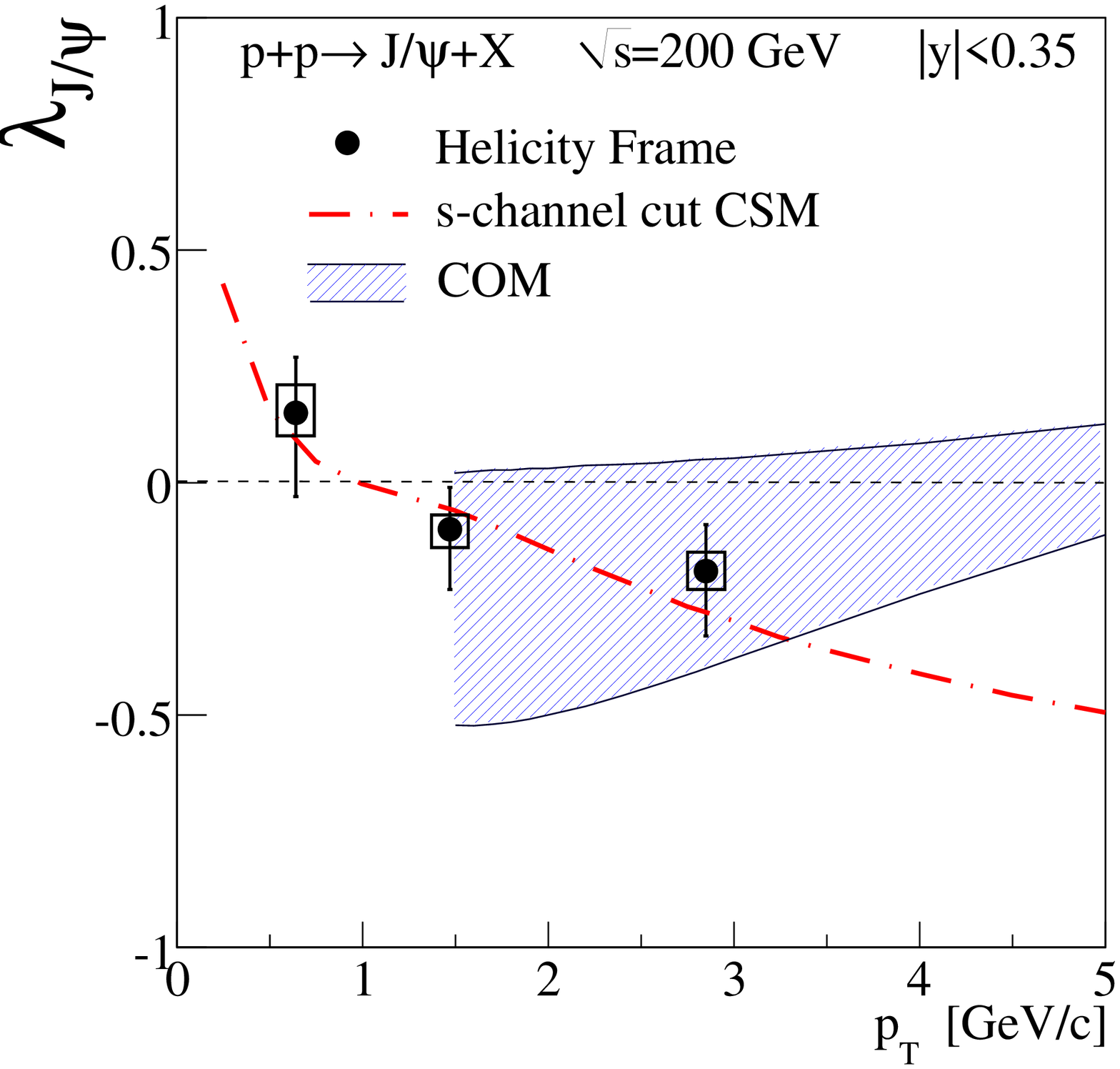}\\
    \includegraphics[width=0.9\linewidth]{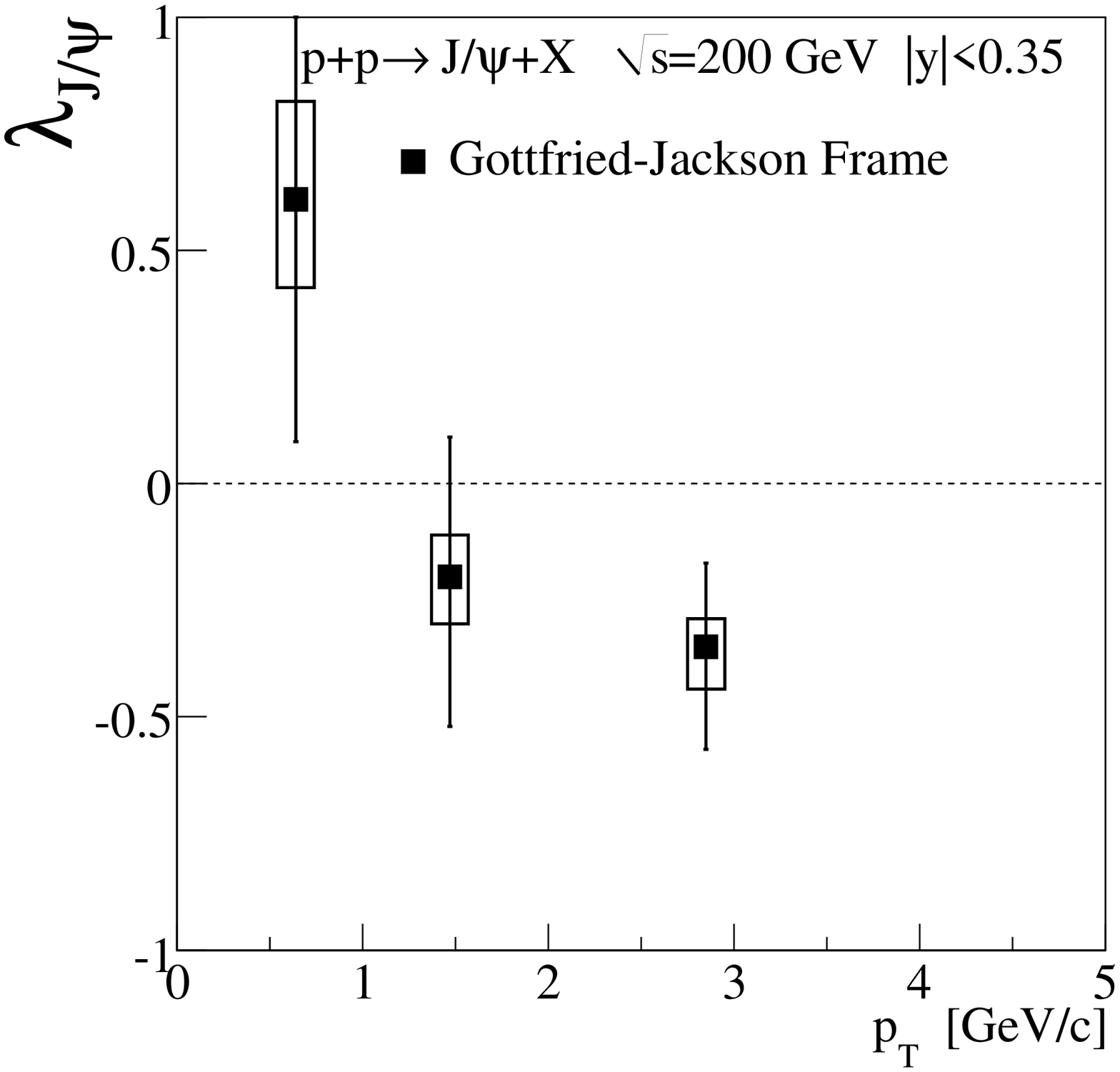}\\
  \caption{\label{pol_pt} \jpsi polarization
parameter $\func{\lambda_{J/\psi}}$ versus transverse
momentum $\func{p_T}$.  Boxes are correlated systematic uncertainties.
(upper) Helicity frame data is compared with
COM\cite{Chung:2009xr} and s-channel CSM \cite{Haberzettl:2007kj}
calculated in the same polarization frame, but there is no prediction
for CEM.  (lower) There are no theoretical predictions
for the Gottfried-Jackson frame.}
\end{figure}

Figure~\ref{pol_pt} shows the transverse momentum dependence of the 
\jpsi polarization in the HX and JG frames.  The uncertainties of 
the fit are larger in the GJ frame given the smaller \costheta 
range compared to HX frame.  The numerical values are listed in 
Table~\ref{tab:results}.  For the HX frame also shown are currently 
available theoretical models:  COM \cite{Chung:2009xr} and the 
s-channel cut CSM \cite{Haberzettl:2007kj} calculated using the 
same polarization frame.  There are no theoretical predictions for 
the GJ frame.

\begin{table}
  \caption{\label{tab:results}\jpsi polarization results in the
    helicity and Gottfried-Jackson frames.  Transverse momentum is in
    \gevc.  Uncertainties correspond to statistical and systematics
    respectively.}
\begin{ruledtabular}
  \begin{tabular}{lccc}
    $p_T~~~$ & \mean{\pt} & $\lambda_{J/\psi}^{HX}$ &
    $\lambda_{J/\psi}^{GJ}$\\ \hline \\
    0-1 & 0.64 & $0.15^{+0.12}_{-0.18}$ $^{+0.06}_{-0.05}$ 
& $0.61^{+0.39}_{-0.52}$ $^{+0.21}_{-0.19}$\\ \\
    1-2 & 1.47 & $-0.10^{+0.09}_{-0.13}$ $^{+0.03}_{-0.04}$ 
& $-0.20^{+0.30}_{-0.32}$ $^{+0.09}_{-0.10}$\\ \\
    2-5 & 2.85 & $-0.19^{+0.10}_{-0.16} \pm 0.04$ 
& $-0.35^{+0.18}_{-0.22}$ $^{+0.06}_{-0.09}$\\ \\
    0-5 & 1.78 & $-0.10^{+0.05}_{-0.09} \pm 0.05$  
& $-0.16^{+0.18}_{-0.12}$ $^{+0.09}_{-0.11}$\\ \\
  \end{tabular}
\end{ruledtabular}
\end{table}

The measurements presented here are for inclusive \jpsi.  Feed-down 
from $\chi_c$ and $\psi^{\prime}$ may also contribute to the observed 
polarization and are not separated out.  The world average result for 
the feed-down contribution to the \jpsi yield is $33 \pm 5\%$ 
\cite{Faccioli:2008ir}.  The polarization of the indirect \jpsi 
should be smeared during the decay process.  If the \jpsi from 
feed-down sources are unpolarized, the direct \jpsi may have a larger 
$\lambda$ in magnitude than that reported here.

\begin{figure}
\centering
  \includegraphics[width=1.0\linewidth]{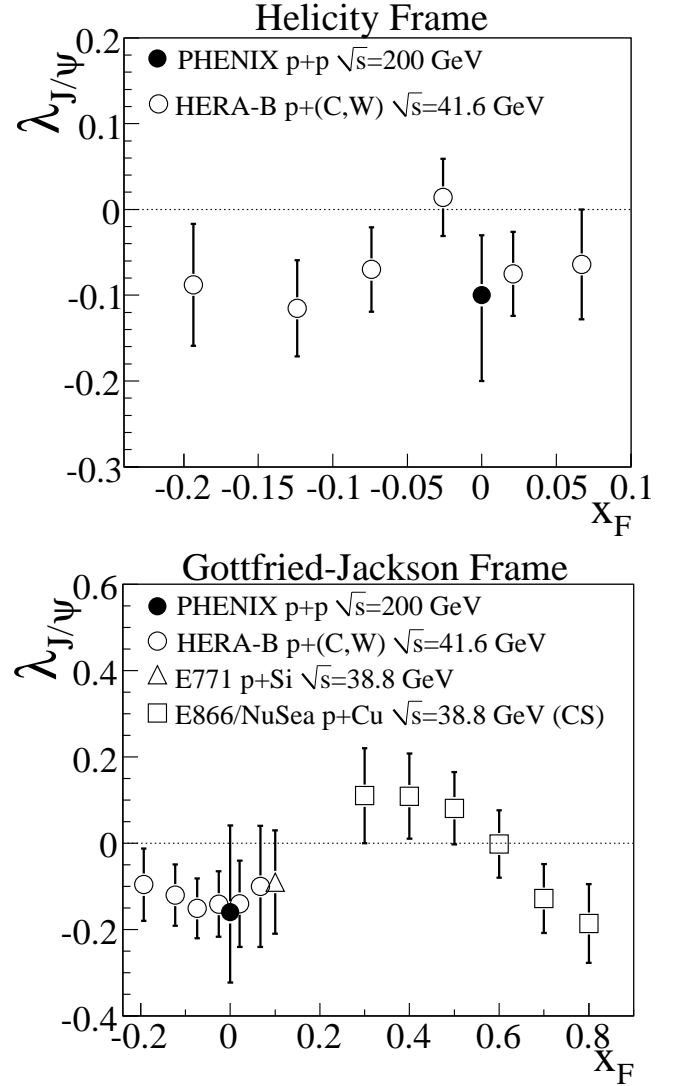}\\
  \caption{\label{fig:pol_xf}$x_F$ dependence of \jpsi polarization for
    $p_T<5~\gevc~$ measured by PHENIX, HERA-B
    \cite{Abt:2009nu}, E771 \cite{Alexopoulos:1997yd} and
    E866/NuSea (CS frame)\cite{Chang:2003rz}.}
\end{figure}

The \jpsi polarization is consistent with zero for all transverse 
momenta but exhibits a 1.8 sigma longitudinal polarization at 
\mbox{$p_T > 2$ \gevc$~$} in the HX and GJ frame when the quadratic 
sum of the statistical and systematic uncertainties are considered.  
In the HX frame, the \pt dependent $\lambda$ follows the s-channel 
cut CSM expectations for prompt \jpsi \cite{Lansberg:2008jn}.  
Finally, the COM prediction \cite{Chung:2009xr}, using the NRQCD 
matrix elements fitted to CDF data, is also consistent with our data 
over the \pt range covered by the calculation.

Figure~\ref{fig:pol_xf} shows that the polarization for
\mbox{$p_T < 5~\gevc~$} follows what is observed in fixed target
experiments for a more extended $x_F$ range in the HX and GJ
frames.  Statistical and systematic uncertainties are
quadratically summed for this comparison.  Note that the
E866/NuSea result was measured in the CS frame.

In principle, intermediate singlet and octet color states may be 
absorbed differently in the nuclear matter present for fixed target 
$p$+A measurements, possibly changing the final \jpsi polarization.  
However, comparisons are limited by the uncertainties in the 
present data.  The observed agreement between the $p+p$ results 
reported here and fixed target $p$~+~A measurements are not yet 
able to determine the magnitude of nuclear matter effects on \jpsi 
polarization.  Direct comparison between future high statistics \pt 
and rapidity dependence of the \jpsi polarization in \pp and $d$+Au 
collisions will provide a better picture for these effects.

\section{Conclusions}

We have presented the first \jpsi polarization measurement at RHIC 
for two different polarization frames.  The observed \pt-dependent 
\jpsi polarization parameter in the HX frame is consistent with the 
s-channel cut CSM, COM and no polarization within current 
uncertainties.  The integrated momentum polarization observed in 
both the HX and GJ frames are in good agreement with the results 
obtained at fixed target experiments collected in lower energy $p$+A 
collision in the same $x_F$ region.  Upcoming higher luminosity \pp 
data will allow more accurate measurements over the full decay 
angular distributions and over extended \pt and rapidity ranges.

\section*{Acknowledgments}
\label{sec:ack}


We thank the staff of the Collider-Accelerator and Physics
Departments at Brookhaven National Laboratory and the staff of
the other PHENIX participating institutions for their vital
contributions.  We acknowledge support from the 
Office of Nuclear Physics in the
Office of Science of the Department of Energy,
the National Science Foundation, 
a sponsored research grant from Renaissance Technologies LLC, 
Abilene Christian University Research Council, 
Research Foundation of SUNY, 
and Dean of the College of Arts and Sciences, Vanderbilt University (USA),
Ministry of Education, Culture, Sports, Science, and Technology
and the Japan Society for the Promotion of Science (Japan),
Conselho Nacional de Desenvolvimento Cient\'{\i}fico e
Tecnol{\'o}gico and Funda\c c{\~a}o de Amparo {\`a} Pesquisa do
Estado de S{\~a}o Paulo (Brazil),
Natural Science Foundation of China (People's Republic of China),
Ministry of Education, Youth and Sports (Czech Republic),
Centre National de la Recherche Scientifique, Commissariat
{\`a} l'{\'E}nergie Atomique, and Institut National de Physique
Nucl{\'e}aire et de Physique des Particules (France),
Ministry of Industry, Science and Tekhnologies,
Bundesministerium f\"ur Bildung und Forschung, Deutscher Akademischer 
Austausch Dienst, and Alexander von Humboldt Stiftung (Germany),
Hungarian National Science Fund, OTKA (Hungary), 
Department of Atomic Energy (India), 
Israel Science Foundation (Israel), 
National Research Foundation (Korea),
Ministry of Education and Science, Russia Academy of Sciences,
Federal Agency of Atomic Energy (Russia),
VR and the Wallenberg Foundation (Sweden), 
the U.S.  Civilian Research and Development Foundation for the
Independent States of the Former Soviet Union, 
the US-Hungarian Fulbright Foundation for Educational Exchange,
and the US-Israel Binational Science Foundation.



\end{document}